\documentclass[final,5p,times,compress,twocolumn]{elsarticle}

\usepackage{amsmath,amssymb}
\usepackage{lipsum}
\usepackage{graphicx}% Include figure files
\usepackage{dcolumn}% Align table columns on decimal point
\usepackage{bm}% bold math
\usepackage{hyperref}% add hypertext capabilities
\usepackage[caption=false]{subfig}
\usepackage[usenames,dvipsnames]{xcolor}
\usepackage{color}
\usepackage{cancel}
\usepackage[normalem]{ulem}

\newcommand{\msun}{$M_\odot$}

\journal{Physics Letters B}

\begin{document}

\begin{frontmatter}

\title{ Unraveling the global behavior of equation of state \\
by explicit finite nuclei constraints}

\author[first]{Anagh Venneti}
\ead{p20210060@hyderabad.bits-pilani.ac.in}
\author[first,second]{Sakshi Gautam}
\author[first]{Sarmistha Banik}
\author[third,fourth]{B. K. Agrawal}
\affiliation[first]{organization={Department of Physics, BITS-Pilani Hyderabad campus},%Department and Organization
            city={Hyderabad},
            postcode={500078}, 
%           state={Telangana},
            country={India}}
\affiliation[second]{organization={Department of Physics, Panjab University},%Department and Organization
            city={Chandigarh},
            postcode={160014}, 
            country={India}}
\affiliation[third]{organization={Saha Institute of Nuclear Physics},%Department and Organization
            addressline={1/AF Bidhannagar}, 
            city={Kolkata},
            postcode={700064}, 
            state={West Bengal},
            country={India}}
\affiliation[fourth]{organization={Homi Bhabha National Institute},%Department and Organization
            addressline={Anushakti Nagar}, 
            city={Mumbai},
            postcode={400094}, 
            state={Maharashtra},
            country={India}}

\begin{abstract}

We obtain posterior distribution of equations of state (EOSs) across a broad range of density by imposing explicitly the constraints from precisely measured fundamental properties of finite nuclei, in combination with the experimental data from heavy-ion collisions and the astrophysical observations of radius, tidal deformability and minimum-maximum mass of neutron stars. The acquired EOSs exhibit a distinct global behavior compared to those usually obtained by imposing the finite nuclei constraints implicitly through empirical values of selected key parameters describing symmetric nuclear matter and symmetry energy in the vicinity of the saturation density. The explicit treatment of finite nuclei constraints yields softer EOSs at low densities which eventually become stiffer to meet the maximum mass criteria. The Kullback-Leibler divergence has been used to perform a quantitative comparison of the distributions of neutron star properties resulting from the EOSs obtained from implicit and explicit finite nuclei constraints.

\end{abstract}


\end{frontmatter}

\noindent
\textbf{Introduction} \\
\label{introduction}
The wealth of data \cite{Sorensen2024} stemming from nuclear physics experiments corresponding to fundamental properties of finite nuclei and heavy-ion collisions (HICs) together with astrophysical observations has spurred comprehensive investigations aimed at constraining the equation of state (EOS) for neutron star matter over a wide range of density. The nuclear physics experiments, determining the binding energies \cite{Wang_2021}, charge radii \cite{angeli2013table}, neutron skin thickness \cite{Adhikari2021}, giant resonances \cite{Dutra2012}, and other properties of the finite nuclei, probe the EOS in the vicinity of the nuclear saturation density ($\rho_0 \sim 0.16$ fm$^{-3}$). The HIC experiments \cite{Danielewicz2002,Fevre2016} also give deep insights into the EOS of symmetric nuclear matter and symmetry energy at densities up to 2$\rho_0$. On the other hand, neutron stars (NS) serve as a unique gateway into the exploration of neutron-rich nuclear matter EOS at densities not yet possible in terrestrial experiments. The EOS of dense nuclear matter significantly shapes the properties of NS, such as mass, radius, and tidal deformability. These properties have been observed in pulsar mass measurements \cite{fonseca2021refined}, simultaneous mass and radius measurements from NICER \cite{riley2019nicer,riley2021nicer,miller2019psr,miller2021radius}, and detection of gravitational waves from binary neutron star (BNS) merger events \cite{abbott2018gw170817rad}.
The third-generation gravitational-wave observatories, Cosmic Explorer
and Einstein Telescope, are expected to provide more accurate constraints on the
neutron star equation of state \cite{walker2024precision}.
                                                         
Several efforts have been made to constrain the EOS for neutron star matter combining the results from finite nuclei, heavy-ion collisions as well as astrophysical observations \cite{Zhang_2018,Tsang2020,huth2022constraining}. The constraints from finite nuclei properties have been incorporated implicitly through empirical values of selected nuclear matter parameters (NMPs). These parameters are instrumental in governing the EOS for symmetric nuclear matter and the symmetry energy, around the saturation density, which are the main ingredients for our understanding of the EOS for neutron star matter. Of late, the posterior distribution of EOSs was obtained using a comprehensive set of data from nuclear physics experiments and recent astrophysical observations. The inputs from nuclear physics implicitly included the constraints, on the symmetry energy and symmetry energy pressure, empirically derived using the nuclear masses, neutron skin thickness and iso-vector giant dipole resonances. The experimental data on the HICs were considered in terms of the symmetric nuclear matter pressure and the symmetry energy in the range 0.2-2$\rho_0$ \cite{tsang2023determination}.

Often, the constraints from finite nuclei are incorporated implicitly through the empirical values of binding energy per nucleon, incompressibility coefficient for the symmetric nuclear matter, and the symmetry energy, at the saturation density, determined by accurately 
 calibrating the mean-field models to the experimental data on binding energies, charge radii, and iso-scalar giant monopole resonance (ISGMR) in several nuclei. Their values are found to vary over narrow ranges across various models. Consequently, the values of binding energy per nucleon and saturation density are usually kept fixed around $-16$ MeV and $0.16$ fm$^{-3}$, respectively. One usually overlooks the small variations required in these quantities specific to a given model, which may subsequently influence the values of symmetry energy parameters. The uncertainty in the symmetry energy parameters may propagate to NS properties such as radius and tidal deformability \cite{Zhang_2018,Carriere_2003, Patra2023_2,imam2023direct,Lattimer_2001}. It would be ideal to allow all the NMPs to vary and be explicitly subjected to the constraints on the experimental and astrophysical observations, narrowing the parameter space and preserving their inter-correlations. \cite{Malik2020,Patra2023,Malik_2022}.

The precisely measured values of finite nuclei properties, such as binding energies, charge radii, and giant resonances, when considered explicitly, impose strong constraints on the radius and tidal deformability of NS with canonical mass (1.4 \msun). The radius $R_{1.4} = 10.6 - 12.6$ km and dimensionless tidal deformability $\Lambda_{1.4} = 125 - 410$ were estimated solely from the constraints imposed by the finite nuclei properties and minimum-maximum mass of NS \cite{Malik2019}. Similar analysis with further inclusion of low-density EOS for pure neutron matter from ab-initio calculations yielded $R_{1.4} = 12.3 - 12.5$ km and $\Lambda_{1.4} = 380 - 460$ \cite{Tsang2019}. Recently, the connection between finite nuclei properties and NS with canonical mass has been investigated by employing a group of more than 400 mean field models, derived from the Skyrme-type effective nucleon-nucleon interactions and the effective Lagrangian describing nucleon interactions through $\sigma$, $\omega$, and $\rho$ mesons \cite{Carlson2023}. The analysis reveals that the models adept at capturing the binding energies, charge radii and ISGMR energies of symmetric and asymmetric finite nuclei display a strong correlation between tidal deformability of canonical mass neutron star and the slope of symmetry energy at the saturation density.

In our present work, we obtain a posterior distribution of EOSs for neutron star matter, consistent with the measured values of basic properties of finite nuclei, such as binding energies and charge radii, together with the comprehensive set of experimental data from HICs and astrophysical observations of radius, tidal deformability and minimum-maximum mass of neutron stars. We compare these EOSs with their counterpart obtained by implicitly imposing the finite nuclei constraints. The statistical distance between the distributions of EOSs resulting from these two distinct scenarios is quantified using the Kullback-Leibler divergence. Additionally, the distributions of radius and tidal deformability across a wide range of neutron star masses are also calculated. 

\noindent
\textbf{Methodology} \\
 The nuclear matter is widely described by the formalism of relativistic mean field (RMF) theory, based on quantum hadrodynamics. In the RMF theory, the neutrons and protons, constituents of the nuclear matter, interact through the exchange of $\sigma$, $\omega$, and $\rho$ mesons. The $\sigma$ and $\omega$ mesons respectively mediate the medium-range attractive and short-range repulsive nature of nuclear interactions, respectively, while the $\rho$ meson governs the iso-vector properties. Additionally, the theory incorporates cross interaction between $\omega$ and $\rho$ mesons as well as the non-linear self-couplings of the $\sigma$ and $\omega$ mesons. The Lagrangian of this nuclear matter is given as \cite{Dutra2014,Zhu_2023},
 \begin{equation}\label{eqn:1}
   \mathcal{L}_{NL} = \mathcal{L}_{nm} + \mathcal{L}_{\sigma} + \mathcal{L}_{\omega} + \mathcal{L}_{\rho} + \mathcal{L}_{int},
\end{equation}
where,
\begin{eqnarray}
    \mathcal{L}_{nm}&=& \!\bar{\psi}\left(i\gamma^\mu\partial_\mu - m\right)\psi +\!\!g_\sigma \sigma \bar{\psi} \psi -\!\!g_\omega \bar{\psi}\gamma^\mu\omega_\mu\psi 
    \nonumber \\
     &&-\frac{g_\rho}{2}\bar{\psi}\gamma^\mu\vec{\rho}_\mu \vec{\tau}\psi,\nonumber \\
%\begin{eqnarray}
    \mathcal{L}_\sigma&=&\frac{1}{2}\left(\partial^\mu\sigma\partial_\mu\sigma - m^2_\sigma \sigma^2\right) - \frac{A}{3} \sigma^3 - \frac{B}{4} \sigma^4 \nonumber,\\
%\end{eqnarray}
%\begin{eqnarray}
    \mathcal{L}_\omega&=&\! -\frac{1}{4} \Omega^{\mu\nu}\Omega_{\mu\nu} + \frac{1}{2}m^2_\omega\omega^\mu\omega_\mu +\frac{C}{4}\left(g_\omega^2 \omega_\mu \omega^\mu\right)^2 \nonumber,\\
%\end{eqnarray}
%\begin{eqnarray}
    \mathcal{L}_\rho&=& -\frac{1}{4}\vec{B}^{\mu\nu}\vec{B}_{\mu\nu} + \frac{1}{2}m^2_\rho \vec{\rho}_\mu \vec{\rho}^{ \mu} \nonumber, \\
%\end{eqnarray}
%and
%\begin{eqnarray}
    \mathcal{L}_{int}&=&\frac{1}{2} \Lambda_v g^2_\omega g^2_\rho \omega_\mu \omega^\mu \vec{\rho}_\mu \vec{\rho}^\mu. \nonumber
\end{eqnarray}
In the above equations, $\Omega_{\mu\nu} = \partial_\nu \omega_\mu - \partial_\mu \omega_\nu$ and $\vec{B}_{\mu\nu} = \partial_\nu \vec{\rho}_\mu - \partial_\mu \vec{\rho}_\nu - g_\rho \left(\vec{\rho}_\mu \times \vec{\rho}_\nu \right)$. The masses of the nucleon, $\sigma$, $\omega$ and $\rho$ mesons are denoted by $m$, $m_\sigma$, $m_\omega$, and $m_\rho$ respectively. The coupling strengths $g_\sigma$, $g_\omega$, $g_\rho$, $A$, $B$, $C$ and $\Lambda_v$ are usually calibrated to yield the measured values of the bulk properties of finite nuclei and neutron stars accurately. The finite nuclei properties are calculated by numerically solving the equations of motion for the mesons and nucleons within the basis expansion method \cite{Gambhir1989,Gambhir1990, RING1997}. 

The properties of a static, non-rotating neutron star (NS), such as mass and radius, can be determined by a pair of coupled differential equations called the Tolman-Oppenheimer-Volkoff (TOV) equations \cite{Tolman1939,Oppenheimer1939, glendenning2012compact}. The dimensionless tidal deformability (in the natural units $G=c=1$) , defined as $\Lambda = \frac{2}{3} k_2 \left(\frac{R}{M}\right)^5$, $M$ and $R$ being the mass and radius, is a measure of the tendency of the NS to deform under the tidal forces. The tidal love number ($k_2$) can be computed by solving a differential equation along with the TOV equations \cite{Hinderer2008}. These equations require the EOS for beta-equilibrated charge-neutral matter. The EOS for crust is constructed from the Baym-Pethick-Sutherland \cite{baym1971ground} EOS and a polytropic fit \cite{Carriere_2003,Malik2017}. The EOS of the core, which starts at 0.5 $\rho_0$, can be readily obtained for the given Lagrangian (Eq.\ref{eqn:1}) \cite{Dutra2014}.
\begin{table*}%[htb]%The best place to locate the table environment is directly after its first reference in text
\begin{center}
\caption{\label{tab:FN_prop_table}%
The median and 68\% CI for binding energies (B) and charge radii ($R_{ch}$) for a few selected nuclei were obtained using implicit and explicit FNCs. The `*' highlights the experimental data \cite{Wang_2021,angeli2013table} for the nuclei used for the explicit FNC.}
\bgroup
\def\arraystretch{1.5}
\begin{tabular}{p{1cm}p{2cm}p{1.5cm}p{2cm}p{1.5cm}p{1.5cm}p{1.25cm}}
\hline

&\multicolumn{2}{c}{Implicit-FNC}&\multicolumn{2}{c}{Explicit-FNC}&\multicolumn{2}{c}{Experimental}\\
\hline
&B [MeV]&$R_{ch}$ [fm]&B [MeV]&$R_{ch}$ [fm]&B [MeV]&$R_{ch}$ [fm]\\
\hline
\hline
$^{16}$O&-135.85$^{+8.91}_{-9.29}$&2.62$^{+0.03}_{-0.03}$&-129.4$^{+1.14}_{-1.16}$&2.7$^{+0.01}_{-0.01}$&-127.62&2.70\\
$^{40}$Ca$^*$&-353.87$^{+19.29}_{-19.64}$&3.34$^{+0.05}_{-0.03}$&-342.73$^{+1.63}_{-1.69}$&3.44$^{+0.01}_{-0.01}$&-342.05&3.48\\
$^{48}$Ca&-420.24$^{+22.28}_{-22.51}$&3.38$^{+0.06}_{-0.03}$&-414.39$^{+1.87}_{-2.2}$&3.47$^{+0.01}_{-0.02}$&-416.00&3.48\\
$^{132}$Sn&-1096.68$^{+56.56}_{-56.22}$&4.6$^{+0.1}_{-0.05}$&-1103.84$^{+5.64}_{-5.49}$&4.72$^{+0.02}_{-0.02}$&-1102.84&4.71\\
$^{208}$Pb$^*$&-1620.19$^{+88.37}_{-87.82}$&5.38$^{+0.12}_{-0.06}$&-1632.94$^{+7.73}_{-7.66}$&5.53$^{+0.02}_{-0.02}$&-1636.43&5.50\\
\hline
\hline
\end{tabular}
\egroup
\end{center}
\end{table*}

We follow the Bayesian approach in our statistical analysis of the EOSs. Bayesian analysis utilizes the Bayes' theorem, defined as,
\begin{equation}
    P(\bm{\Theta}|D) = \frac{\mathcal{L}(D|\bm{\Theta})P(\bm{\Theta})}{\mathcal{Z}},
\end{equation}
where the $\mathcal{L}(D|\bm{\Theta})$, $P(\bm{\Theta})$, and $P(\bm{\Theta}|D)$ are the likelihood function, the prior and the posterior for the nuclear matter parameters $\Theta = \{\rho_0, E_0, K_0, Q_0, m^*/m, J_0, L_0\}$, given the data $D$. The parameters in $\Theta$ are used to determine the seven coupling parameters $g_\sigma$, $g_\omega$, $g_\rho$, $A$, $B$, $C$, and $\Lambda_v$ (as elaborated in Section A of the supplemental material (SM)). These coupling parameters, in turn, are used to determine finite nuclei properties, the EOS, and subsequently, NS properties. The priors and posteriors are listed in Table S1 of the SM. We use a Gaussian likelihood, defined as,
\begin{equation}\label{gauss}
\mathcal{L}(D|\bm{\Theta}) = \prod_j \frac{1}{\sqrt{2\pi\sigma^2_j}} e^{-\frac{1}{2}\left(\frac{d_j - m_j(\Theta)}{\sigma_j}\right)^2},
\end{equation}
where $m_j$, $d_j$, and $\sigma_j$ are the predicted value for a given set of the model parameters, the observed value, and the uncertainty respectively. The index $j$ runs over the experimental and astrophysical observations. To incorporate the observational constraints of tidal deformability \cite{abbott2018gw170817rad}, we use an asymmetric Gaussian \cite{Tsang2020,tsang2023determination}. The Bayesian analysis is conducted utilizing Python libraries of BILBY \cite{Ashton2019} and pymultinest \cite{buchner2014x}. \\

\noindent
\textbf{Results and Discussions} \\
Following Tsang \textit{et al} \cite{tsang2023determination}, we place the constraints on the symmetry energy $J(\rho)$ (Eq.S4 of SM), pressure of symmetric nuclear matter $P_{SNM}(\rho)$ (Eq.S5 of SM), the symmetry pressure $P_{sym}(\rho)$ (Eq.S6 of SM), and the NS observables such as radius, tidal deformability, and minimum-maximum mass. The analyses of dipole polarizability of $^{208}$Pb \cite{Zhang2015}, and isospin diffusion data and single ratios of neutron-proton spectra \cite{Tsang2009,Morfouace2019} provide us with empirical constraints on $J(\rho)$ at densities $<0.5\rho_0$. At densities $0.5-1.5 \rho_0$, analyses of nuclear masses \cite{Brown2014,Kortelainen2012,DANIELEWICZ2017}, PREX-II \cite{Adhikari2021}, and pion spectra from heavy ion collisions (HICs) \cite{Lynch2022} have placed constraints on $J(\rho)$ and $P_{sym}(\rho)$. The ISGMR \cite{Dutra2012} places the limits on the incompressibility coefficient ($K_0$). $P_{SNM}(\rho)$ and $P_{sym}(\rho)$ are constrained by HICs at the densities $2\rho_0$ and $\sim 1.5\rho_0$. The astrophysical observations on the minimum-maximum mass \cite{fonseca2021refined}, radius \cite{riley2019nicer,riley2021nicer,miller2019psr,miller2021radius}, and tidal deformability \cite{abbott2018gw170817rad} of NS place constraints on the EOS for densities beyond $2\rho_0$. We label the set of all the aforementioned constraints as implicit finite nuclei constraints (FNC).

We then consider the scenario in which the EOSs are subjected to direct experimental constraints from binding energies \cite{Wang_2021} and charge radii \cite{angeli2013table} for $^{40}$Ca and $^{208}$Pb, which can be computed from the Lagrangian (Eq.\ref{eqn:1}) \cite{Gambhir1989,Gambhir1990}. We incorporate these in place of the constraints from nuclear masses in the case of implicit FNC. This set of constraints is labeled as explicit FNC. We present a comprehensive overview of all the constraints, along with the corresponding median and 68\% confidence interval (CI) obtained from posterior distributions for both scenarios, in Tables S3 and S4 of the SM.\\

\begin{figure}
    \centering
    \includegraphics[width=0.425\textwidth,height=0.27\textwidth]{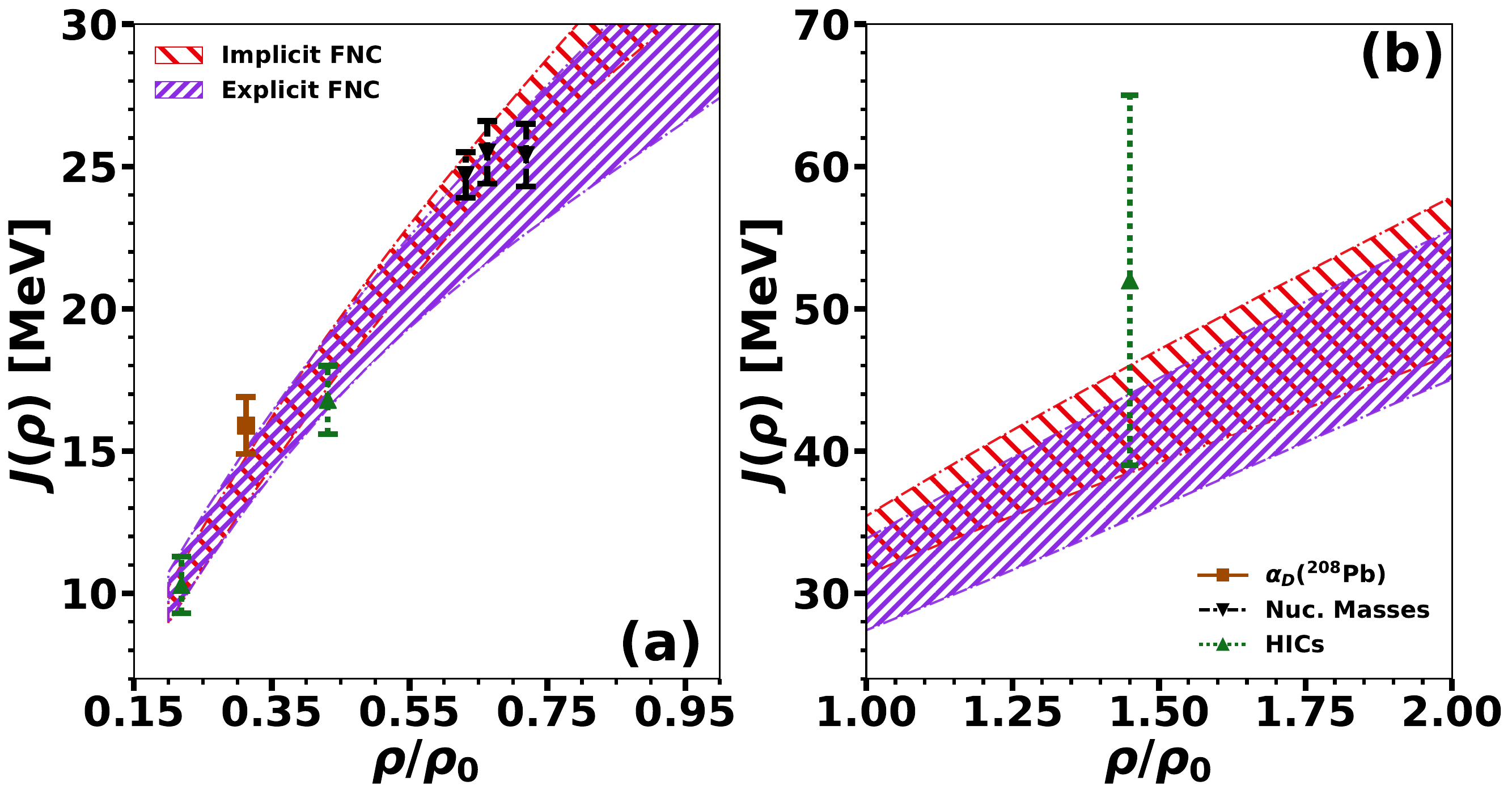}
    \caption{The 68\% CIs of symmetry energy, for the cases of implicit (red) and explicit (violet) FNCs, in the density ranges between (a) 0.15 - 1$\rho_0$ and (b) 1 - 2$\rho_0$. The field symbols and the vertical lines indicate the medians and 68\% CIs of constraints from dipole polarizability (brown), nuclear masses (black), and HICs (dark green).}
    \label{fig:Symmetry_Energy}
\end{figure}
\noindent
\textit{Properties of finite nuclei} \\
In Table \ref{tab:FN_prop_table}, we present the median and 68\% CI of binding energies and charge rms radii, for some selected nuclei, obtained using the posterior distributions of coupling parameters (listed in Table S2 of SM) resulting from the implicit and explicit FNCs. The explicit FNC yields the binding energies and charge radii closer to the corresponding measured values. It is evident that the implicit FNC yields over-binding in the case of lighter nuclei and under-binding for the heavier nuclei. The implicit FNC also yields smaller charge rms radii. The deviation in medians of binding energies ranges from about 6\% for $^{16}$O nucleus to about 1\% for $^{208}$Pb. On the contrary, employing explicit FNC has led to a notable reduction in the deviation, about 1.5\% for $^{16}$O, and less than 0.5\% for all other nuclei under consideration. \\

\begin{figure*}[htb]
\begin{minipage}{0.5\linewidth}
\centering
\includegraphics[width=0.75\linewidth,height=0.7\linewidth,]{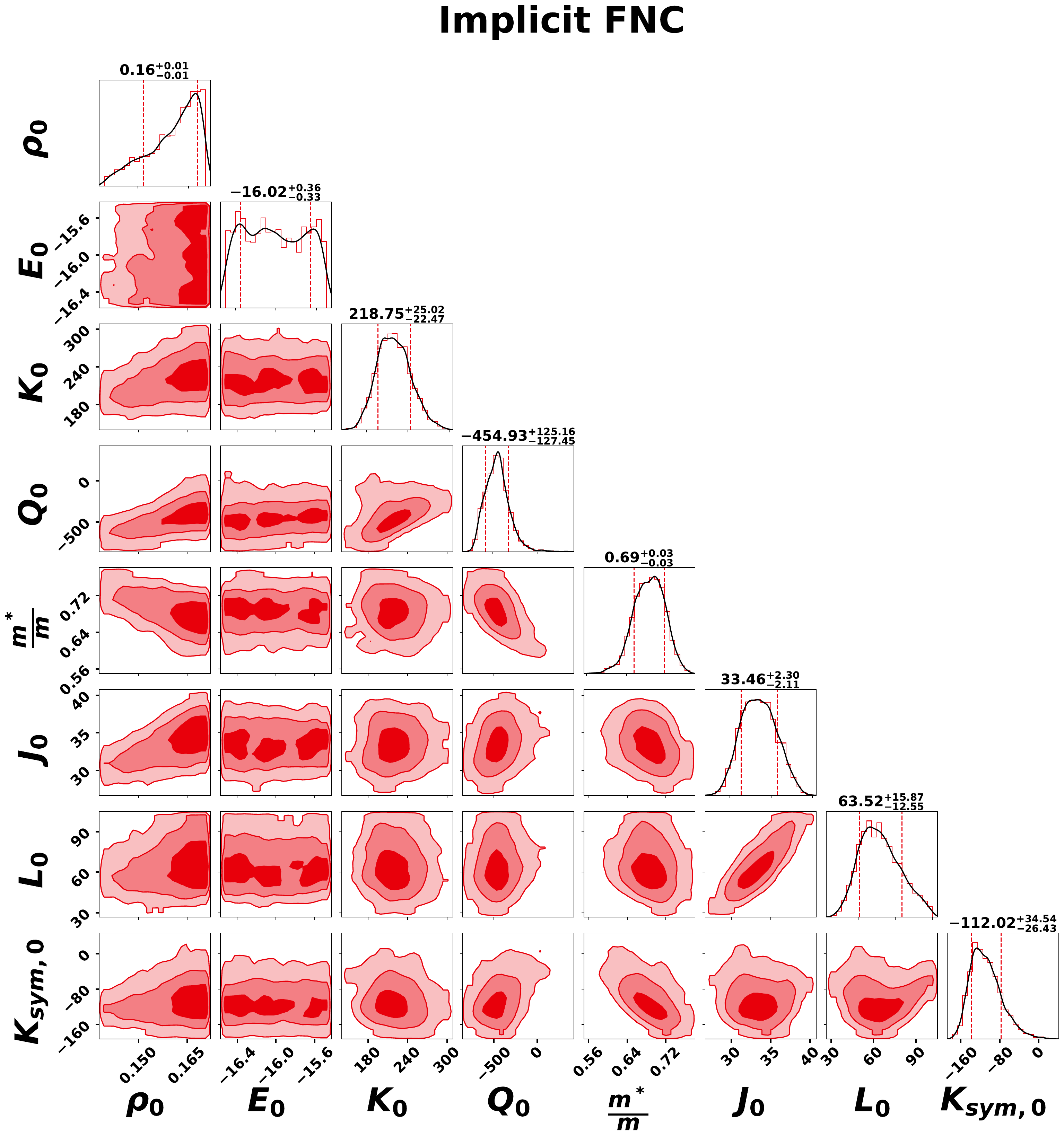} %Final_Plots/Case_1_NMPs_Posterior_Corner_plot.pdf
\label{fig1:sub1}
%\caption{(a)}
\end{minipage}%
\begin{minipage}{0.5\linewidth}
\centering
\includegraphics[width=0.75\linewidth,height=0.7\linewidth,]{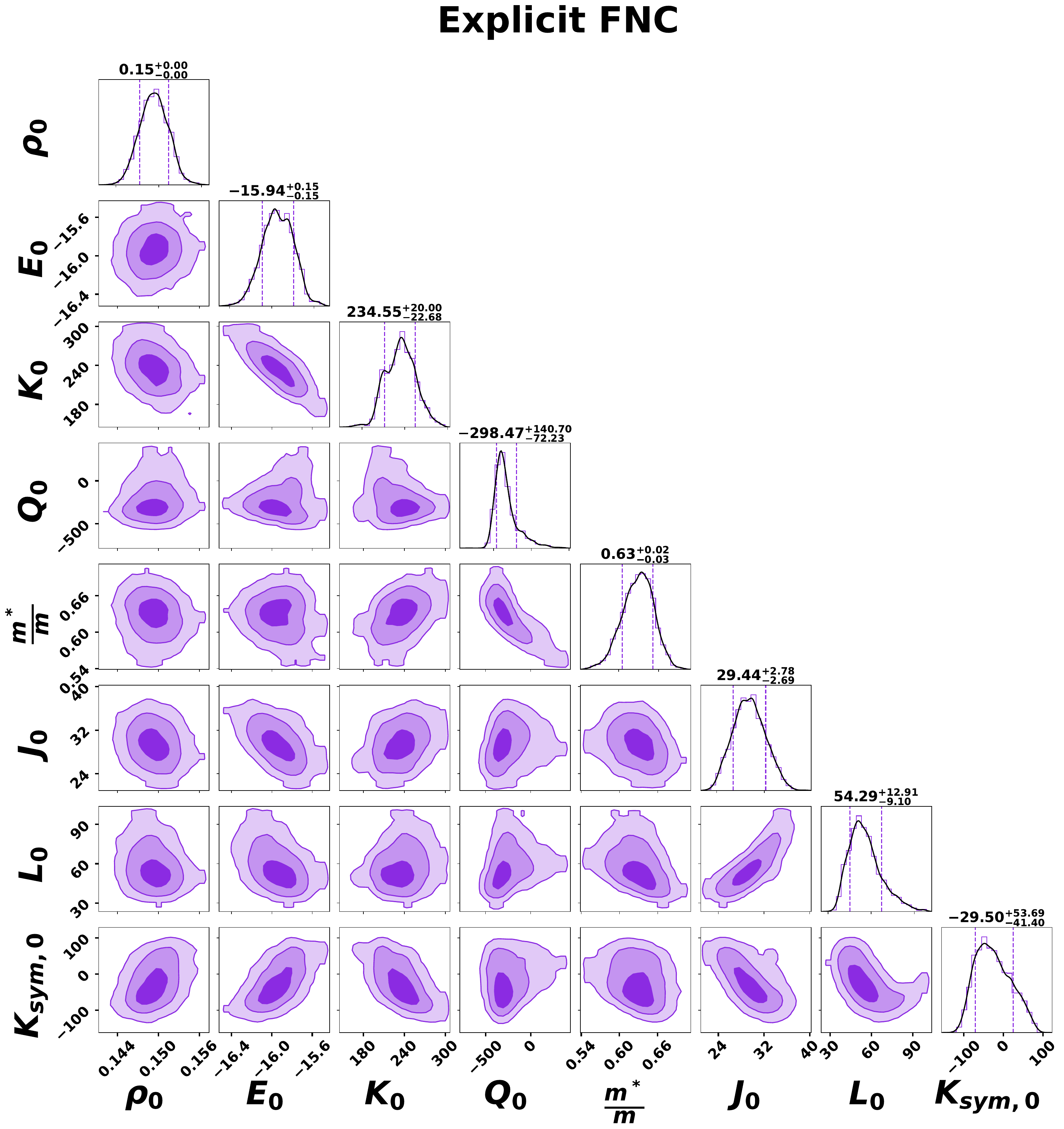}
\label{fig1:sub2}
%\caption{(a)}
\end{minipage}
\caption{Posterior distributions of nuclear matter parameters (NMPs) for the two different scenarios of implicit and explicit FNC. The 1$\sigma$ CI is displayed as vertical dashed lines in the marginalized posterior distributions of the NMPs. All are in the units of MeV except for the saturation density $\rho_0$ (in fm$^{-3}$) and $m^*/m$ (dimensionless).}
\label{post_nmps}
\end{figure*}
\noindent
\textit{Symmetry Energy} \\
Figs. \ref{fig:Symmetry_Energy}a and \ref{fig:Symmetry_Energy}b show 68\% CI for the symmetry energy across the density ranges of 0.15 - 1$\rho_0$ and 1 - 2$\rho_0$, respectively. The vertical dashed lines represent the empirical constraints imposed on the symmetry energy at different densities. 
The symmetry energy, $J(\rho)$, for the Lagrangian (Eq. \ref{eqn:1}),  is as following, \cite{Dutra2014} 
\begin{equation}
    J\left(\rho\right) = \frac{k^2_f}{6 E^*_f} + \frac{1}{8}\left(\frac{g_\rho}{m^{*}_\rho}\right)^2 \rho,
\end{equation}
where $E^*_f = \left(k^2_f + m^{*2}\right)^{1/2}$, and $\left(m^{*}_\rho\right)^2 = m^2_\rho + \Lambda_v g^2_\omega g^2_\rho \omega^2_0$ for Fermi momentum $k_f$, effective nucleon mass $m^* = m - g_\sigma \sigma$ and effective $\rho$-meson mass $m^{*}_\rho$.
Evidently, the symmetry energy distribution resulting from explicit FNC exhibits softer behavior at densities $> 0.5 \rho_0$, in contrast to the behavior of those from the implicit FNC. This distinction arises from incorporating the limits on binding energies and charge radii explicitly in place of those on $E_0$, $\rho_0$, and $J(\rho)$ as employed in the case of implicit FNC. The softness in symmetry energy in the case of explicit FNC persists beyond 1.5$\rho_0$.
Nevertheless, the two scenarios are in similar agreement with respect to the constraints from HICs and dipole polarizability at densities 0.2 - 0.55$\rho_0$. The medians and 68\% CIs of the symmetry energy along with those of $P_{SNM}(\rho)$, $K_0$ and $P_{sym}(\rho)$ obtained from the posterior distribution of EOSs for both the scenarios are presented in Table S3 of the SM. \\

\noindent
\textit{Posterior Distribution of NMPs} \\
The corner plots illustrating the marginalized posterior distributions for both scenarios are presented in Figs.\ref{post_nmps}. On the diagonals, the marginalized posterior distributions for individual NMPs are depicted with vertical dashed lines representing the 1$\sigma$ interval. The lower off-diagonal plots display pairwise probability distributions with contours enclosing the 1$\sigma$, 2$\sigma$, and 3$\sigma$ CIs. The NMPs are better constrained for the explicit FNC scenario, that exhibits a preference for lower values of the parameters, with the exception of the incompressibility coefficient $K_0$, skewness coefficient $Q_0$ and the curvature of symmetry energy $K_{sym,0}$, when compared to the implicit FNC scenario. A NS of mass $\geq 2.08M_\odot$ \cite{fonseca2021refined} requires a stiff EOS at high densities. Consequently, the higher values of $K_0$, $Q_0$, and $K_{sym,0}$ are needed to counterbalance the softness in the symmetry energy observed for the case of explicit constraints (Fig.\ref{fig:Symmetry_Energy}).

The explicit FNC exhibit more inter-correlations, which are eventually stronger, in comparison to those for the implicit FNC. It may be noted that $J_0 -K_{sym,0}$, $L_0 -K_{sym,0}$, $K_0 -E_0$, among other correlations, show significant improvements in the explicit-FNC despite a narrowing down of the variation in the parameters. $K_{sym,0}$ develops stronger correlations with other parameters for the case of explicit FNC. For instance, the correlation of $K_{sym,0}$ with $J_0$ becomes stronger (Pearson correlation coefficient $r\sim-0.7$) in comparison to the negligible correlation observed in the case of implicit FNC. Additionally, the nature of the correlation between $K_0$ and $\rho_0$ transforms from a positive correlation ($r\sim0.41$) in the implicit FNC to an anti-correlation ($r\sim-0.35$) in the explicit FNC. The pairwise Pearson correlation coefficients ($r$) among the NMPs, and the properties of the finite nuclei are shown in the Figs.S1 and S2 of the SM for both the scenarios.\\

\begin{figure}
    \centering
    \includegraphics[width=0.425\textwidth,height=0.25\textwidth]{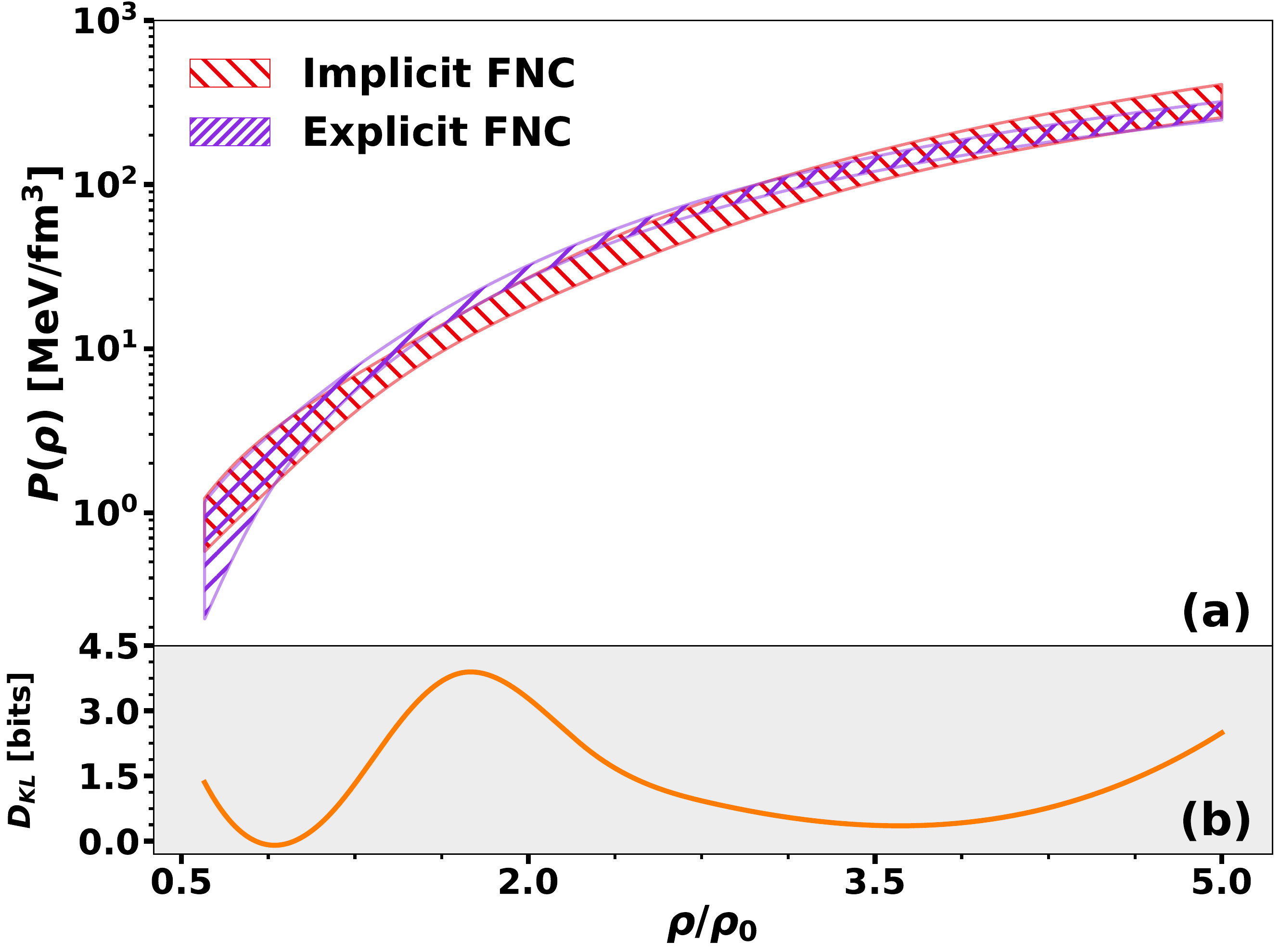}
    \caption{(a) Plots for EOSs displaying a 95\% CI for the scenarios with implicit (red hatch) and explicit (violet hatch) finite nuclei constraints.(b) The Kullback-Leibler divergence (D$_{KL}$ in bits) was obtained using the distributions of EOSs for the two scenarios.}
    \label{fig:EOS}
\end{figure}
\noindent
\textit{EOS and NS properties} \\
Fig.\ref{fig:EOS}a shows the 95\% CI for the posterior distributions of EOSs resulting from explicit and implicit FNCs. In general, the explicit FNC displays a narrower spread as compared to the implicit FNC. The two scenarios display distinct behaviors at different densities. At low densities ($<\rho_0$), the EOSs generated for the case of explicit constraints are softer than those for the implicit constraints. The EOSs obtained from the former exhibit significant stiffness at intermediate densities ($\sim$1.25-2.5$\rho_0$) in comparison to those from the latter. At higher densities ($>$2.5$\rho_0$), the distribution of the EOSs soften once more, overlapping with the distribution of EOSs from implicit constraints.

In Fig.\ref{fig:EOS}b, we plot the values of Kullback-Leibler (KL) divergence ($D_{KL}$) \cite{Kullback1951,mackay2003information}, for a quantitative comparison of the distributions of the EOSs for the two different scenarios considered. The $D_{KL}$ is calculated as,
\begin{eqnarray}
    D_{KL} \left(P||Q\right) = \int^\infty_{-\infty} p(x) log_2\left(\frac{p(x)}{q(x)}\right) dx, \nonumber
\end{eqnarray}
where P and Q represent the probability distributions for the pressure at a given number density for explicit FNC and implicit FNC scenarios, respectively. The statistical distance $D_{KL}$ quantifies the difference between the distributions of the pressure at a given number density, for the two different scenarios considered. The divergence, $D_{KL}$, approaches 0 when the distributions $p(x)$ and $q(x)$ are similar to each other.
The value of $D_{KL}$ is observed to vary across the density range. The values of KL divergence is significant at the intermediate densities (around $\sim 2\rho_0$), indicating noticeable differences between the two distributions. It approaches a minimum value at low ($\sim 0.8\rho_0$) and high densities ($2.75 - 4.25\rho_0$), indicating an evident overlap of these two distributions at these densities (Fig.\ref{fig:EOS}a).
\begin{figure}
	\centering 
	\includegraphics[width=0.37\textwidth,height=0.3\textwidth]{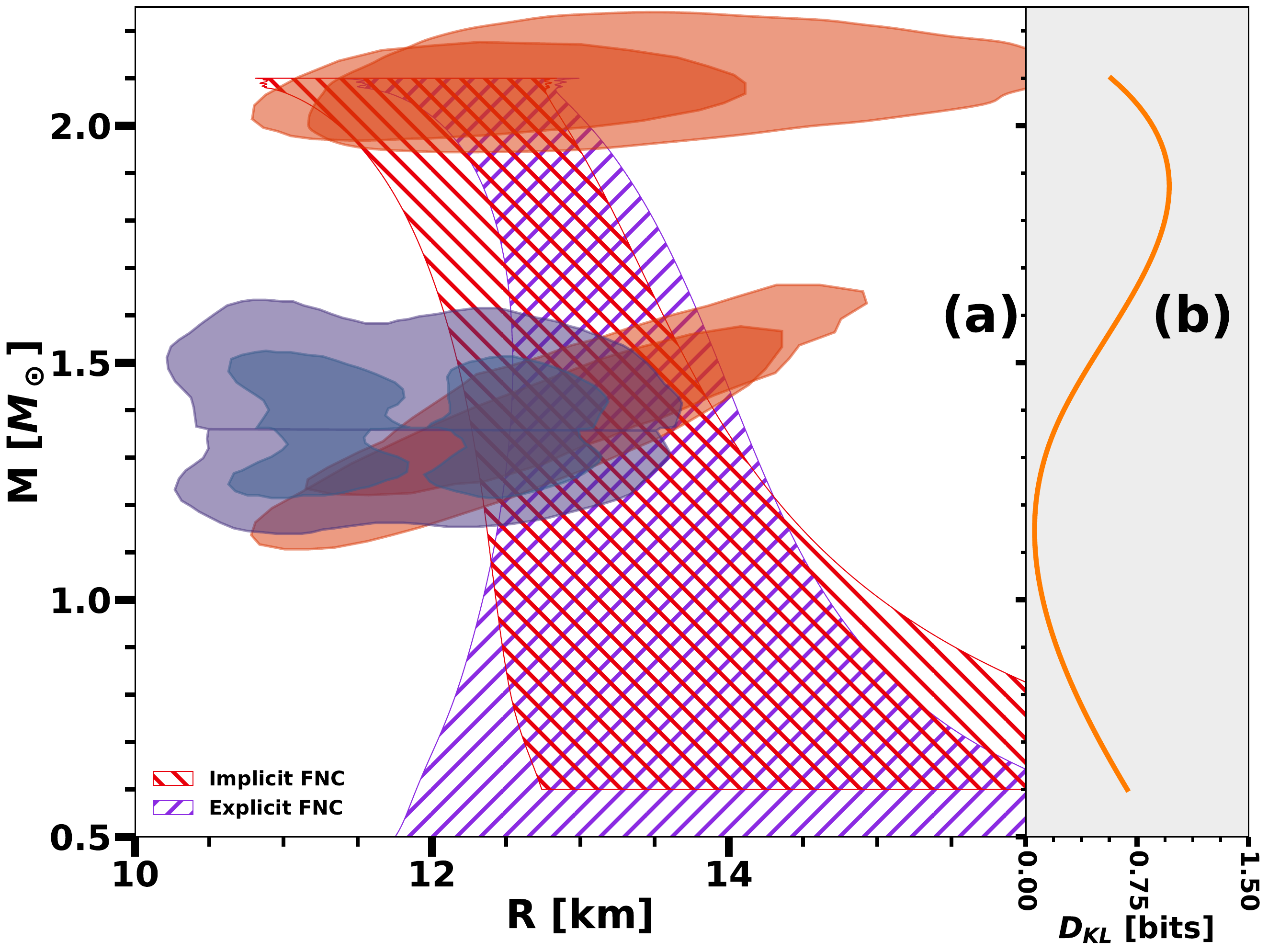}	
	\caption{(a) The 95\% CI neutron star mass-radius plots with implicit (red) and explicit (violet) finite nuclei constraints. The radius inferred from GW170817 (dark blue contours) 
 \citep{abbott2018gw170817rad}, NICER observations (orange contours) for PSR J0030+0451 ($\sim 1.44 M_\odot$) \citep{riley2019nicer,miller2019psr} and PSR J0740+6620 ($\sim 2.08 M_\odot$)\citep{riley2021nicer,miller2021radius} are also shown. (b) The Kullback-Leibler divergence (D$_{KL}$ in bits) was obtained using the distributions of mass-radius for the two scenarios.} % {\sout{for the scenarios}}{\sout{for the scenarios}} 
	\label{fig_mass_radius}%
\end{figure}
\begin{figure}
	\centering 
	\includegraphics[width=0.37\textwidth,height=0.3\textwidth]{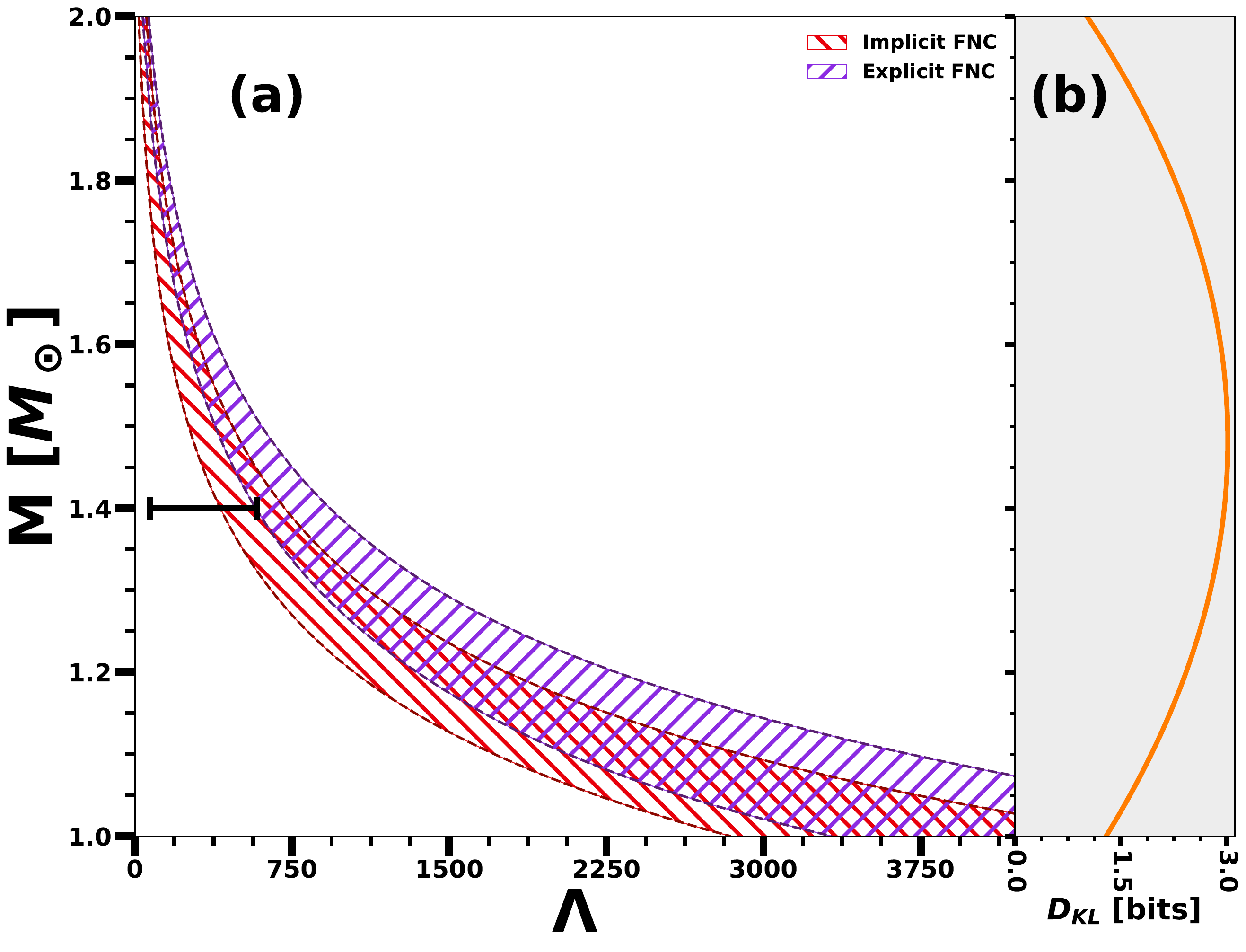}	
	\caption{(a) The 95\% CI neutron star mass-tidal deformability with implicit (red hatch) and explicit (violet hatch) constraints. The canonical tidal deformability from GW170817\citep{abbott2018gw170817rad} is shown (horizontal black line). (b) The Kullback-Leibler divergence (D$_{KL}$ in bits) was obtained using the distributions of mass-tidal deformability for the two scenarios.} % {\sout{for the scenarios}}{\sout{for the scenarios}} 
	\label{fig_tidal}%
\end{figure}

The 95 \% CI of the posterior distributions of mass-radius of NS are plotted in Fig.\ref{fig_mass_radius}a for both the implicit and explicit scenarios. The mass-radius distributions corresponding to the EOSs from both explicit and implicit FNCs exhibit distinct behavior across different mass ranges. Particularly, in the case of explicit FNC, the median values of posterior radius distributions for intermediate ($\sim 1.4 M_\odot$ \cite{riley2019nicer,miller2019psr}) and high-mass ($\sim 2.08 M_\odot$ \cite{riley2021nicer,miller2021radius}) NSs (as listed in Table S4 of the SM) tend to be larger by $\sim 0.4$ km. In Fig.\ref{fig_mass_radius}b, the KL divergence between the radius distributions of explicit and implicit FNCs at different NS masses is shown. It varies across the mass range, reaching a minimum value at $\sim 1.0 M_\odot$. At both low and high masses, the KL divergence reaches a value of $\sim 0.75$ bits. However, despite these relatively low KL divergence values, the median values of the radius distributions of the explicit and implicit scenarios show a significant distinction for NSs of these masses.
% Additionally, the explicit scenario predicts a smaller median radius for low mass ($<1.0M_\odot$) NSs.

In Fig.\ref{fig_tidal}a, the 95\% CIs for dimensionless tidal deformability as a function of NS mass are presented. The observational constraint (90\% CI \cite{abbott2018gw170817rad}) on the canonical tidal deformability ($\Lambda_{1.4}$) aligns more closely with the posteriors of implicit FNC than those of the explicit FNC. The values of KL divergence, shown in Fig.\ref{fig_tidal}b, quantify the differences between the distributions of tidal deformability for the two scenarios. The divergence reaches a maximum value ($\sim 3$bits) at $\sim 1.5 M_\odot$, indicating significant differences between the distributions. This difference is less for NSs with high and low masses, signifying an overlap of the distributions of tidal deformability in these mass ranges. The posterior medians and 68\% CIs of NS radius and tidal deformability, for both cases, are listed in Table S4 of the SM.

\begin{figure*}
  \begin{minipage}{0.5\linewidth}
    \centering
    \includegraphics[width=0.6\linewidth,height=0.6\linewidth]{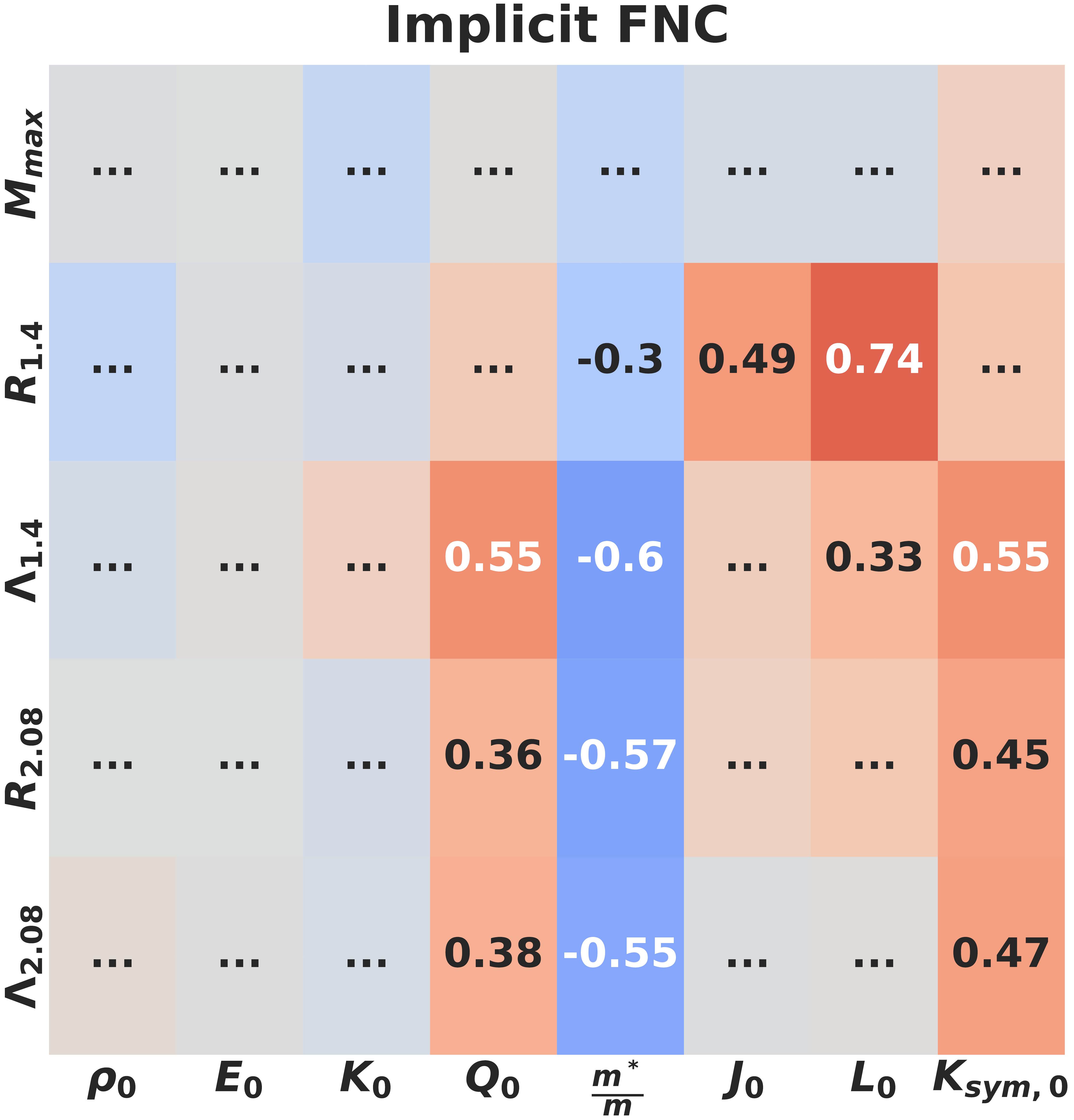}
    \label{fig2:sub1}
  \end{minipage}%
  \begin{minipage}{0.5\linewidth}
    \centering
    \includegraphics[width=0.6\linewidth,height=0.6\linewidth]{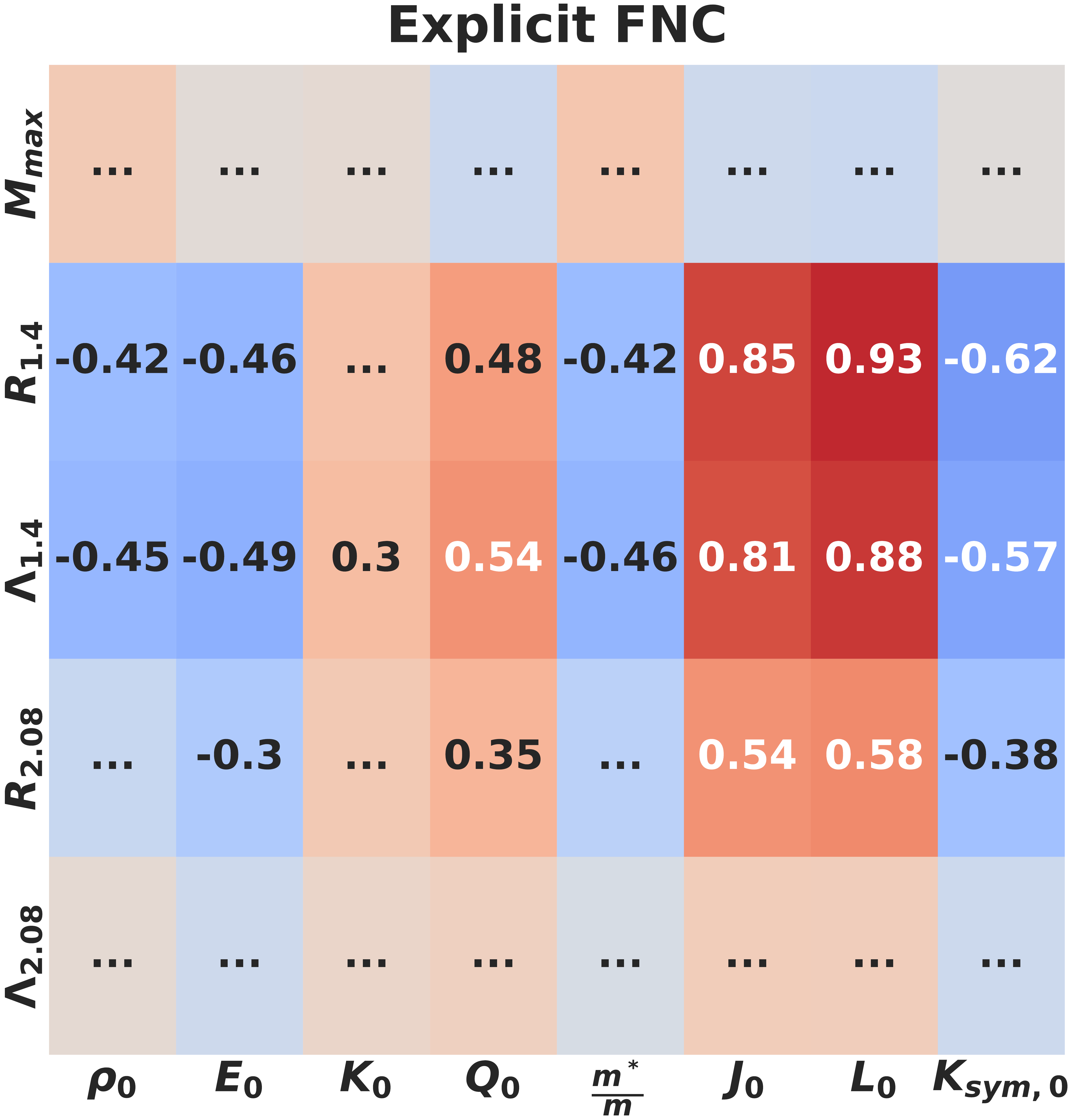}
    \label{fig2:sub2}
  \end{minipage}
  \caption{The correlation heat map between the nuclear matter parameters and the neutron star properties for implicit finite nuclei constraints and explicit finite nuclei constraints. The correlations with absolute values $>0.3$ are displayed. The negative (positive) correlation is shown in the boxes with shades of blue (red), with darker shades indicating a higher negative (positive) correlation.}
  \label{corr_plot}
\end{figure*}
Finally, in Fig.\ref{corr_plot}, the values of Pearson correlation coefficients ($r$) between nuclear matter parameters and neutron star properties for both scenarios are presented. The explicit FNC produces stronger correlations among a larger number of pairs of NMPs and NS properties. The correlations involving parameters $J_0$ and $L_0$, among others, show significant improvement. For instance, $J_0$ - $\Lambda_{1.4}$ correlation is strong ($r\sim0.8$) for explicit FNC as compared to the negligible correlation in the case of implicit FNC. Additionally, $E_0$ and $\rho_0$ show weak correlations with NS properties, which were absent in the case of implicit FNC. Interestingly, the signs of the correlations in some cases are opposite for these two scenarios. E.g., the $K_{sym,0}$ - $\Lambda_{1.4}$ correlation. The stronger correlations as depicted by the explicit FNC scenario
would enable one to place tighter bounds on the values of some of the NMPs from the precise astrophysical observations of the NS properties. 

We have also repeated our analysis for the RMF model without the inclusion of vector self-coupling , i.e. $C=0$. The trends are qualitatively almost similar to those obtained for the model with the $\omega$-meson self-coupling. Regardless of the model under consideration, the explicit constraints yields EOSs that are stiffer at intermediate densities (around $\sim 2\rho_0$) than those obtained for implicit constraints. However, the stiffness of the EOSs persists at higher densities for the model without the vector self coupling (Fig S3 of the SM). Intermediate mass ($\sim 1.44 M_\odot$) NSs, for the model without vector self-coupling, exhibit smaller radius (by $\sim 0.5$ km) as compared to the model incorporating the vector self-coupling. The radius of higher NS mass ($\sim 2.08 M_\odot$), on the other hand, is predicted to be higher by the same amount. The results for the model without the vector self-coupling is presented in section 3 of the SM. Nevertheless, irrespective of the RMF model under consideration, explicit constraints on finite nuclei properties, imposed at low densities, supplemented with astrophysical constraints at high densities exhibit a global effect on the EOS and NS properties. \\
\noindent
\textbf{Summary and outlook} \\
We have confined the distributions of EOSs, derived from the
relativistic mean-field model, under two distinct scenarios regarding finite nuclei constraints. One of the scenarios explicitly considers precisely measured binding energies and charge radii for $^{40}$Ca and $^{208}$Pb nuclei, along with the experimental data from heavy-ion collisions, other finite nuclei observables and astrophysical observations of neutron stars, to establish posterior distributions for the EOSs. We compared our results with the commonly employed scenario in which finite nuclei constraints are imposed implicitly through specific properties describing symmetric nuclear matter and the density dependence of symmetry energy at sub-saturation densities. The two different scenarios yield differing distributions of the NMPs as well as the corresponding EOSs. The correlations among NMPs and NS properties are found to be stronger in the case of explicit scenario. The distributions of EOSs for the two different scenarios considered are employed to obtain the NS properties. A quantitative comparison of the distributions of the EOSs and the NS properties is performed through the Kullback-Leibler (KL) divergence. 

It is observed that the implicit constraints significantly over bind the light nuclei (up to $\sim7\%$) and under bind the heavier nuclei (by $\sim1\%$) while also yielding smaller charge radii. The deviations, from the experimentally measured values, in the case of explicit treatment of finite nuclei constraints are significantly smaller. The EOSs resulting from it exhibit considerable stiffness at intermediate densities ($1.25 - 2.5 \rho_0$). These features are quantitatively reflected in the values of KL divergence, which is significant at these densities. The properties of NSs are also significantly affected. The median values of the radius distributions for explicit and implicit scenarios, at higher NS mass ($\sim 2.08 M_\odot$) and intermediate NS mass ($\sim 1.4 M_\odot$),  differ by $\sim 0.4$ km. This difference is reflected at these masses by the values of KL divergence, which also indicate similar differences at lower masses. The distinction between the scenarios is further highlighted in the posterior distribution of tidal deformability. The explicit FNC also predicts higher canonical tidal deformability ($\Lambda_{1.4}$), as compared to the implicit FNC, displaying a smaller overlap with the observational constraints on $\Lambda_{1.4}$ \cite{abbott2018gw170817rad}. The maximum value of the KL divergence around $\sim 1.4 M_\odot$ emphasizes this difference between the two approaches.

The present investigation may be extended to include other finite nuclei observables, such as iso-scalar and iso-vector giant resonances, explicitly for a diverse set of nuclei to constraint the EOS.\\
\noindent
\textbf{Acknowledgements} \\
AV would like to acknowledge the CSIR for the support through the CSIR-JRF 09/1026(16303)/2023-EMR-I.

\bibliographystyle{elsarticle-num}

\bibliography{References.bib}

\end{document}

\begin{frontmatter}



\title{Supplemental Material : Unraveling the global behavior of equation of state \\
by explicit finite nuclei constraints}


\author[first]{Anagh Venneti}
\author[first,second]{Sakshi Gautam}
\author[first]{Sarmistha Banik}
\author[third,fourth]{Bijay K. Agrawal}
\affiliation[first]{organization={Department of Physics, BITS-Pilani Hyderabad campus},%Department and Organization %           addressline={}, 
            city={Hyderabad},
            postcode={500078}, 
            state={Telangana},
            country={India}}
\affiliation[second]{organization={Department of Physics, Panjab University},%Department and Organization
%           addressline={}, 
            city={Chandigarh},
            postcode={160014}, 
            state={Panjab},
            country={India}}
\affiliation[third]{organization={Saha Institute of Nuclear Physics},%Department and Organization
            addressline={1/AF Bidhannagar}, 
            city={Kolkata},
            postcode={700064}, 
            state={West Bengal},
            country={India}}
\affiliation[fourth]{organization={Homi Bhabha National Institute},%Department and Organization
            addressline={Anushakti Nagar}, 
            city={Mumbai},
            postcode={400094}, 
            state={Maharasthra},
            country={India}}

\end{frontmatter}
\appendix
\renewcommand{\thesection}{A}
\renewcommand{\theequation}{S\arabic{equation}}
\section{The Model}
The energy density can be computed, at a given number density ($\rho = \rho_p + \rho_n$; $\rho_{p/n}$ is the proton/neutron number density), for the Lagrangian (Eq.1 of the main paper) as, \cite{Dutra2014}

\begin{eqnarray}\label{eden}
    \mathcal{E}(\rho, y)&=&\frac{1}{2} m^2_\sigma \sigma^2 + \frac{A}{3} \sigma^3 + \frac{B}{4} \sigma^4 - \frac{1}{2} m^2_\omega \omega^2_0 
     -\frac{C}{4} \left(g^2_\omega \omega^2_0\right)^2- \frac{1}{2} m^2_\rho \bar{\rho}^2_{0(3)} + g_\omega \omega_0 \rho + \frac{g_\rho}{2} \bar{\rho}_{0(3)} \left(2 y - 1\right)\rho  
     \\
     & & -\frac{1}{2} \Lambda_\nu g^2_\omega g^2_\rho \omega^2_0 \bar{\rho}^2_{0(3)} + \sum_{i=p,n}\mathcal{E}^{i}_{kin} \nonumber \\
     P(\rho,y) &=& \rho^2 \frac{\partial (\mathcal{E}(\rho,y)/\rho)}{\partial \rho} \label{press}
\end{eqnarray}
where the proton fraction $y = \rho_p/\rho$. The kinetic energy contributions to energy density, $\mathcal{E}^{i}_{kin}$ ($i=p/n$ for protons/neutrons), is, %$\rho = (\rho_p + \rho_n)$,
\begin{eqnarray}
    & \mathcal{E}^{i}_{kin} = \frac{\gamma k^4_{F i}}{2 \pi^2} \left[\left(1 + \frac{z^2} {2}\right)\frac{\sqrt{1 + z^2}}{4} - \frac{z^4}{8}ln\left(\frac{1 + \sqrt{1+z^2}}{z}\right)\right]
\end{eqnarray}
%   \: \text{and}
% & P^{p/n}_{kin} = \frac{\gamma k^4_{F p/n}}{6 \pi^2} \left[\left(1 - \frac{3z^2} {2}\right)\frac{\sqrt{1 + z^2}}{4} + \frac{3z^4}{8} ln\left(\frac{1 + \sqrt{1+z^2}}{z}\right)\right]. \nonumber
Here,  $z = m^*/k_{F i}$,  $m^* = m - g_\sigma \sigma$ and $k_{F i}$ being the effective nucleon mass and the Fermi momentum respectively, and $\gamma = 2$ corresponds to spin degeneracy. The coupling parameters (in Eq.\ref{eden}), $g_\sigma$, $g_\omega$, $A$, $B$, and $C$, are determined using iso-scalar nuclear matter parameters (NMPs) while the iso-vector NMPs determine $g_\rho$ and $\Lambda_v$. The iso-scalar parameters considered are the energy per nucleon $E_0$, incompressibility coefficient $K_0$, skewness coefficient $Q_0$ and nucleon effective mass $m^*$ while iso-vector parameters are the symmetry energy $J_0$ and its slope $L_0$, evaluated at the saturation density $\rho_0$ \cite{Dutra2014,Zhu_2023}. These NMPs can be defined as,
\begin{eqnarray}
 E_0 &=& \left(\mathcal{E}(\rho, y)/\rho\right)_{\rho = \rho_0, y = 1/2}, \nonumber \\
 K_0 &=& 9 \left(\frac{\partial P(\rho, y)}{\partial \rho}\right)_{\rho = \rho_0, y = 1/2}, \nonumber \\
 Q_0 &=& 27 \rho^3_0 \left(\frac{\partial^3\left(\mathcal{E}(\rho, y)/\rho\right)}{\partial \rho}\right)_{\rho = \rho_0, y = 1/2}, \\
 J_0 &=& J(\rho = \rho_0) = \frac{1}{8} \left(\frac{\partial^2 (\mathcal{E}(\rho, y)/\rho)}{\partial y^2}\right)_{\rho = \rho_0, y = 1/2}, \text{and} \nonumber\\
 L_0 &=& 3 \rho_0 \left(\frac{\partial J\left(\rho\right)}{\partial \rho}\right)_{\rho = \rho_0}.\nonumber
\end{eqnarray}

% K_{sym} = 9 \rho^2 \left(\frac{\partial^2 J\left(\rho\right)}{\partial\rho^2}\right)
The energy density, $\mathcal{E}(\rho, y)$ (Eq.\ref{eden}), and pressure $P(\rho,y)$(Eq.\ref{press}) are the nucleonic contributions to the equation of state (EOS) for the core region of a neutron star (NS). The leptonic contributions, arising from electrons and muons, are added by treating them as free relativistic particles. The nucleon and lepton fractions are determined by adjusting the proton fraction, $y$, such that the conditions of charge neutrality and beta equilibrium are satisfied. The pressure of symmetric nuclear matter $P_{SNM}(\rho)$ and the symmetry pressure $P_{sym}(\rho)$ are defined as,
\begin{eqnarray}
    P_{SNM}(\rho) &=& \rho^2 \left(\frac{\partial (\mathcal{E}(\rho, y)/\rho)}{\partial \rho}\right)_{y=1/2}, \text{and} \\
    P_{sym}(\rho) &=& \rho^2 \left(\frac{\partial J(\rho)}{\partial \rho}\right).
\end{eqnarray}
We can also define the symmetry energy curvature ($K_{sym,0}$)  at the saturation density $\rho_0$ as,
\begin{eqnarray}
 K_{sym,0} = 9 \rho^2 \left(\frac{\partial^2 J\left(\rho\right)}{\partial\rho^2}\right)_{\rho = \rho_0}
\end{eqnarray}
\renewcommand{\thesection}{B}
\section{Priors, Posteriors and Likelihoods}
The joint posterior distribution of the model parameters are determined by the priors and likelihood function (Eq 3 of the main paper). The priors for these NMPS are listed in Table \ref{tab:table1}, along with the posterior median and 68\% CI for both Implicit and Explicit scenarios. % \sim 0.16 fm$^{-3}$
\renewcommand{\thetable}{S1}
\begin{table}[htb!]%The best place to locate the table environment is directly after its first reference in text
\begin{center}
\caption{\label{tab:table1}%
\av{Uniform priors are defined for the NMPs. All are in the units of MeV except for $\rho_0$ (in fm$^{-3}$) and $m^*/m$ (dimensionless). Also listed are the posterior medians and 68\% CI for all the parameters for implicit and explicit FNCs}}
\bgroup
\def\arraystretch{1.5}
\begin{tabular}{p{2cm}P{1.5cm}P{1.65cm}P{1.5cm}P{2cm}P{1.5cm}P{1.5cm}P{1.5cm}}
\hline
\hline
\multicolumn{8}{c}{Prior}\\
\hline
 &$\rho_0$&$E_0$&$K_0$&$Q_0$&$m^*/m$&$J_0$&$L_0$\\ 
 \hline
Min&0.140&-16.5&150&-1500&0.5&20&20\\
Max&0.170&-15.5&300&400&0.8&40&100\\ 
\hline
\multicolumn{8}{c}{Posterior}\\
\hline
Implicit FNC & 0.1621$^{+0.0056}_{-0.0106}$ & -16.02$^{+0.36}_{-0.33}$ & 218.75$^{+25.02}_{-22.47}$ & -454.93$^{+125.16}_{-127.45}$ & 0.69$^{+0.03}_{-0.03}$ & 33.46$^{+2.30}_{-2.11}$ & 63.52$^{+15.87}_{-12.55}$ \\
Explicit FNC & 0.1494$^{+0.002}_{-0.002}$ & -15.94$^{+0.15}_{-0.15}$ & 234.55$^{+20.00}_{-22.68}$ & -298.47$^{+140.70}_{-72.23}$ & 0.63$^{+0.02}_{-0.03}$ & 29.44$^{+2.78}_{-2.69}$ & 54.29$^{+12.91}_{-9.10}$ \\
\hline
\hline
\end{tabular}
\egroup
\end{center}
\end{table}

\av{The posterior medians and 68\% confidence intervals for the coupling parameters, $g_\sigma$, $g_\omega$, $g_\rho$, $A$, $B$, $C$ and $\Lambda_v$, are also listed in Table \ref{coupling} for both implicit and explicit scenarios.}

\renewcommand{\thetable}{S2}
\begin{table}[hbt!]
\begin{center}
    \caption{\label{coupling}% 
    \av{The posterior medians and 68\% CI of the coupling parameters fitted to the nuclear matter parameters for both implicit and explicit FNCs. The nucleon and meson masses $m$, $m_\sigma$, $m_\omega$, and $m_\rho$ are 939 MeV, 508.1941 MeV, 782.501 MeV and 763.00 MeV respectively.}}
    \bgroup
\def\arraystretch{1.5}
    \begin{tabular}{p{3cm}P{1.5cm}P{1.5cm}P{1.5cm}P{1.5cm}P{1.65cm}P{1.55cm}P{1cm}}
    \hline
    \hline
    Coupling Parameter &$g_\sigma$&$g_\omega$ & $g_\rho$ & A & B & C & $\Lambda_v$\\
    \hline
    Implicit FNC & 9.05$^{+0.29}_{-0.31}$ & 10.69$^{+0.61}_{-0.59}$ & 9.95$^{+0.49}_{-0.48}$ & 15.29$^{+3.99}_{-3.06}$& -31.56$^{+8.87}_{-11.19}$& 0.0006$^{+0.0008}_{-0.0005}$& 0.06$^{+0.03}_{-0.03}$\\
    Explicit FNC & 10.0$^{+0.38}_{-0.3}$ & 12.52$^{+0.67}_{-0.54}$& 10.32$^{+0.75}_{-0.62}$&10.62$^{+1.21}_{-1.38}$& -14.07$^{+6.56}_{-5.52}$& 0.0027$^{+0.0005}_{-0.0006}$& 0.08$^{+0.05}_{-0.03}$\\
    \hline
    \hline
    \end{tabular}
    \egroup
\end{center}
    \end{table}
% We use a Gaussian likelihood, defined as,
% \begin{equation}\label{gauss}
% \mathcal{L}(D|\bm{\Theta}) = \prod_j \frac{1}{\sqrt{2\pi\sigma^2_j}} e^{-\frac{1}{2}\left(\frac{d_j - m_j(\Theta)}{\sigma_j}\right)^2},
% \end{equation}
% where $m_j$, $d_j$, and $\sigma_j$ are the predicted value for a given set of the model parameters, the observed value, and the uncertainty respectively. The index $j$ runs over the experimental and observational data. Some of the observational constraints on NS properties have asymmetric uncertainties. To incorporate these, we use an asymmetric Gaussian \cite{Tsang2020,tsang2023determination}.

% \newpage
\renewcommand{\thetable}{S3}
\begin{table*}[hbt!]%The best place to locate the table environment is directly after its first reference in text
\begin{center}
    \caption{\label{tab:liklelihood_snm}%
    \av{The experimental constraints on the properties $P_{SNM}(\rho)$ (in MeV-fm$^{-3}$), $K_0$ (in MeV), $J(\rho)$ (in MeV) and $P_{sym}(\rho)$ (in MeV-fm$^{-3}$). The posterior medians and 68\% CIs of these properties are also listed for both implicit and explicit FNCs. `*' indicates the constraints, from analyses of nuclear masses, employed in the case of implicit FNC.}}
    \bgroup
\def\arraystretch{1.5}
    %$P_{SNM}(\rho)$ [MeV/fm$^{-3}$] &%
    \begin{tabular}{p{1.25cm}p{2.5cm}p{1.75cm}p{2cm}p{2cm}p{2cm}p{2cm}p{1cm}}
    \hline
    \hline
Quantity& Source &$\rho$ [fm$^{-3}$] & Expt. & Implicit FNC & Explicit FNC & Ref.\\
    \hline 
% \multicolumn{7}{c}{Symmetric Nuclear Matter} \\
\hline
$P_{SNM}(\rho)$ &HIC (DLL)& 0.32& 10.1$\pm$3.0&\multirow{2}{*}{14.0$^{+2.2}_{-1.99}$}&\multirow{2}{*}{20.56$^{+1.32}_{-1.38}$}& \cite{Danielewicz2002}\\
&HIC (FOPI)& 0.32& 10.3$\pm$2.8&&&\cite{Fevre2016}\\
\hline
$K_0$ & ISGMR &0.16& 230$\pm$30&218.75$^{+25.02}_{-22.47}$&234.55$^{+20.0}_{-22.68}$&\cite{Dutra2012} \\
\hline
% \multicolumn{7}{c}{Asymmetric Nuclear Matter} \\
% \hline
$J(\rho)$&$\alpha_D$&0.05&15.9$\pm$1.0&13.97$^{+0.57}_{-0.56}$&14.09$^{+0.65}_{-0.61}$&\cite{Zhang2015}\\
&HIC(Isodiff)&0.035& 10.3$\pm$1.0&10.34$^{+0.48}_{-0.47}$&10.62$^{+0.62}_{-0.55}$&\cite{Lynch2022,Tsang2009}\\
&HIC(n/p ratio)& 0.069 & 16.8$\pm$1.2&18.19$^{+0.55}_{-0.55}$&17.83$^{+0.78}_{-0.69}$&\cite{Lynch2022,Morfouace2019}\\
&HIC($\pi$)& 0.232 & 52$\pm$13 & 41.97$^{+4.0}_{-3.25}$&39.29$^{+4.48}_{-3.71}$&\cite{Lynch2022,Estee2021}\\
&Masses(Skyrme)$^*$& 0.101&24.7$\pm$0.8&24.38$^{+0.47}_{-0.49}$&23.04$^{+1.42}_{-1.52}$&\cite{Lynch2022,Brown2014}\\
&Masses(DFT)$^*$& 0.115&25.4$\pm$1.1&26.79$^{+0.62}_{-0.67}$&24.96$^{+1.85}_{-1.87}$&\cite{Lynch2022,Kortelainen2012}\\
&IAS$^*$& 0.106&25.5$\pm$1.1&25.25$^{+0.51}_{-0.51}$&23.74$^{+1.58}_{-1.63}$&\cite{Lynch2022,DANIELEWICZ2017}\\
\hline
$P_{sym}(\rho)$&PREX-II& 0.11 &2.38$\pm$0.75&2.05 $^{+0.28}_{-0.3}$&1.65 $^{+0.4}_{-0.36}$&\cite{Lynch2022,Adhikari2021,Reed2021} \\
&HIC($\pi$)&0.232 & 10.9$\pm$8.7&6.08 $^{+1.52}_{-1.09}$&6.64 $^{+0.84}_{-0.65}$&\cite{Lynch2022,Estee2021}\\
&HIC(n/p flow)&0.240 & 12.1$\pm$8.4 &6.5 $^{+1.56}_{-1.16}$&7.11 $^{+0.83}_{-0.67}$&\cite{Lynch2022}\\
\hline
\hline
    \end{tabular}
    \egroup
\end{center}
    \end{table*}
The constraints on $P_{SNM}(\rho)$ are obtained from analyses of heavy ion collisions (HICs) data \cite{Danielewicz2002, Fevre2016}. Additionally, the analyses of dipole polarizablity ($\alpha_D$) of Pb$^{208}$ \cite{Zhang2015}, masses of doubly magic nuclei using Skyrme models \cite{Brown2014}, deformed nuclei \cite{Kortelainen2012}, isobaric analog states (IAS) \cite{DANIELEWICZ2017}, neutron skin thickness of Pb$^{208}$ (PREX-II) \cite{Adhikari2021} and HICs \cite{Lynch2022} also impose  restriction on values of $J(\rho)$ and $P_{sym}(\rho)$. We list these in Table \ref{tab:liklelihood_snm} along with the posterior medians and 68\% CIs for both implicit and explicit FNCs.
\renewcommand{\thetable}{S4}
\begin{table}[hbt!]
\begin{center}
    \caption{\label{tab:liklelihood_ns}% 
     \av{The astrophysical observations of neutron star radius (in km) and dimensionless tidal deformability. The posterior medians and 68\% CIs are also listed for both implicit and explicit FNCs. `*' marks the NICER constraints which are given a weight of 0.5}} %lower limit of maximum mass (in \msun),}
    \bgroup
\def\arraystretch{1.5}
    \begin{tabular}{p{2.5cm}p{5cm}p{1.5cm}p{2cm}p{2cm}p{1cm}}
    \hline
    \hline
    Quantity& Source & Obs. & Implicit FNC & Explicit FNC & Ref. \\
    \hline
    % Maximum Mass & PSR J0740+6620 & $\geq 2.08 \pm 0.07$ & 2.15$^{+0.1}_{-0.05}$ & 2.11$^{+0.05}_{-0.02}$&\cite{fonseca2021refined}\\ 
    % \hline
    R (1.34 \msun) & NICER (PSR J0030+0451) Riley$^{*}$ & 12.71$^{+1.14}_{-1.19}$ & 13.12$^{+0.46}_{-0.41}$&13.49$^{+0.6}_{-0.53}$&\cite{riley2019nicer}\\
    R (1.44 \msun) & NICER (PSR J0030+0451) Miller$^{*}$ & 13.02$^{+1.24}_{-1.06}$ & 13.0$^{+0.42}_{-0.39}$&13.42$^{+0.55}_{-0.48}$&\cite{miller2019psr}\\
    R (2.07 \msun) & NICER (PSR J0740+6620) Riley$^{*}$ & 12.39$^{+1.30}_{-0.98}$ & 11.87$^{+0.46}_{-0.46}$&12.29$^{+0.37}_{-0.32}$&\cite{riley2021nicer}\\
    R (2.08 \msun) & NICER (PSR J0740+6620) Miller$^{*}$ & 13.7$^{+2.6}_{-1.5}$ &  11.83$^{+0.47}_{-0.47}$&12.24$^{+0.38}_{-0.32}$&\cite{miller2021radius}\\
    % \hline
    $\Lambda$ (1.4 \msun) & GW170817 &190$^{+390}_{-120}$&557.39$^{+80.05}_{-72.08}$&752.72$^{+85.18}_{-84.53}$&\cite{abbott2018gw170817rad}\\
    \hline
    \hline
    \end{tabular}
    \egroup
\end{center}
    \end{table}

The observations of NS properties, such as mass, radius, and tidal deformability, also impose limits on the EOS. The minimal observational constraint for the NS mass is determined by the mass measurement of PSR J0740+6620, $2.08 \pm 0.07$ \cite{fonseca2021refined}. 
The mass and radius of pulsars PSR J0030+0451 (1.4 \msun) and PSR J0740+6620 (2.08 \msun) have been concurrently measured by the Neutron star Interior Composition ExploreR (NICER), providing additional constraints to further refine the parameter space. However, considering that each of these pulsars has been independently analyzed, we assign a weight of 0.5 to each of these constraints \cite{tsang2023determination}.  We use an asymmetric Gaussian likelihood \cite{tsang2023determination,Tsang2020} to incorporate these constraints on the canonical tidal deformability ($\Lambda_{1.4}$) \cite{abbott2018gw170817rad} . Table \ref{tab:liklelihood_ns} lists the observational constraints, posterior medians, and their 68\% CIs of the NS radius and tidal deformability for both the scenarios considered.
% \newpage
\renewcommand{\thefigure}{S1}
\begin{figure}[!ht]
    \centering
    \includegraphics[width=\textwidth,height=1.05\textwidth]{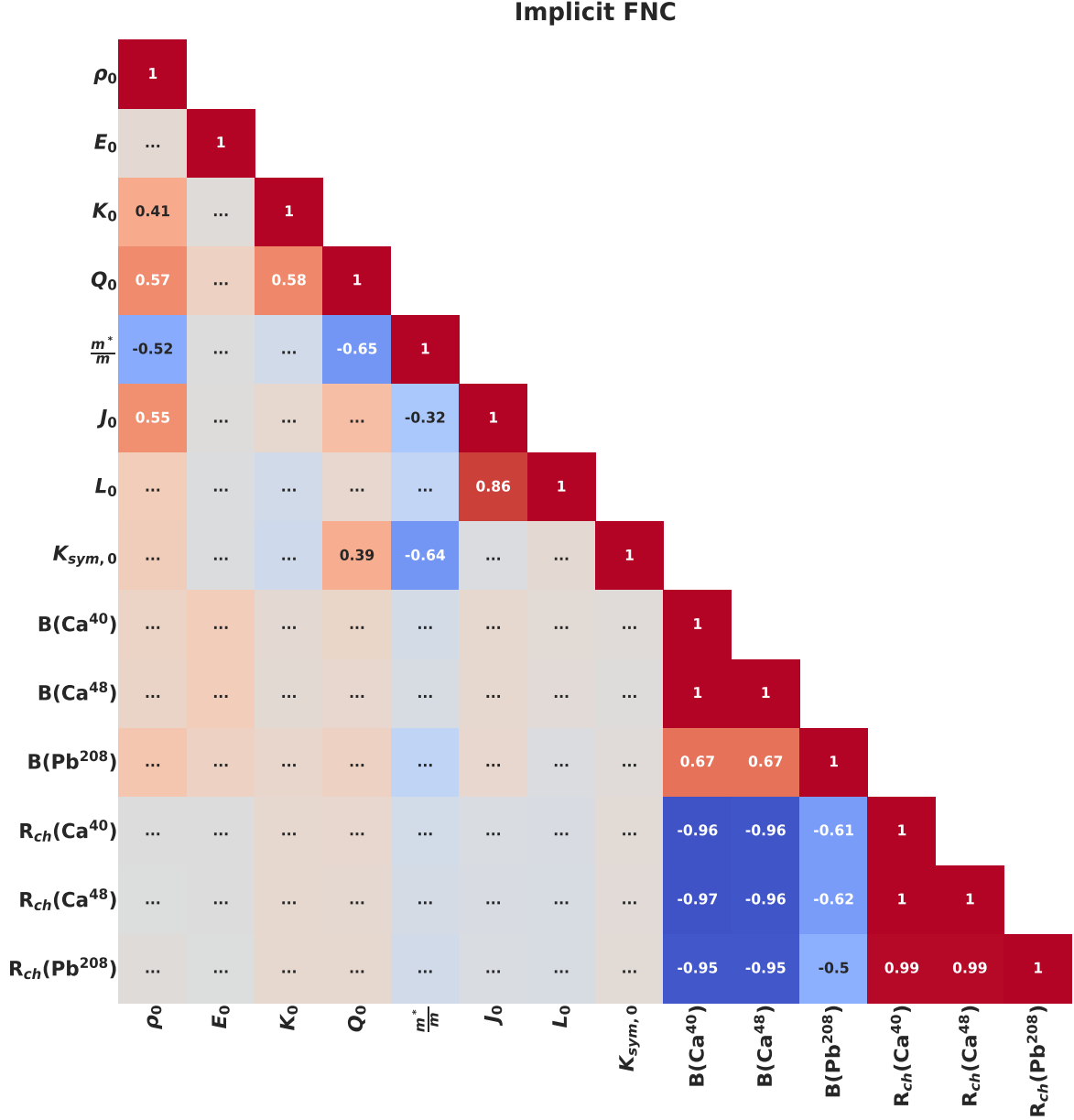}
    \caption{The values of pairwise Pearson correlation coefficients, $|r|>0.3$, of the nuclear matter parameters, binding energies and charge radii of Ca$^{40}$,Ca$^{48}$ and Pb$^{208}$. The negative (positive) correlation is shown in the boxes with shades of blue (red), with darker shade indicating a higher negative (positive) correlation.}
    \label{fig:corr_case1}
\end{figure}
% \newpage
\renewcommand{\thefigure}{S2}
\begin{figure}[!ht]
    \centering
    \includegraphics[width=\textwidth,height=1.05\textwidth]{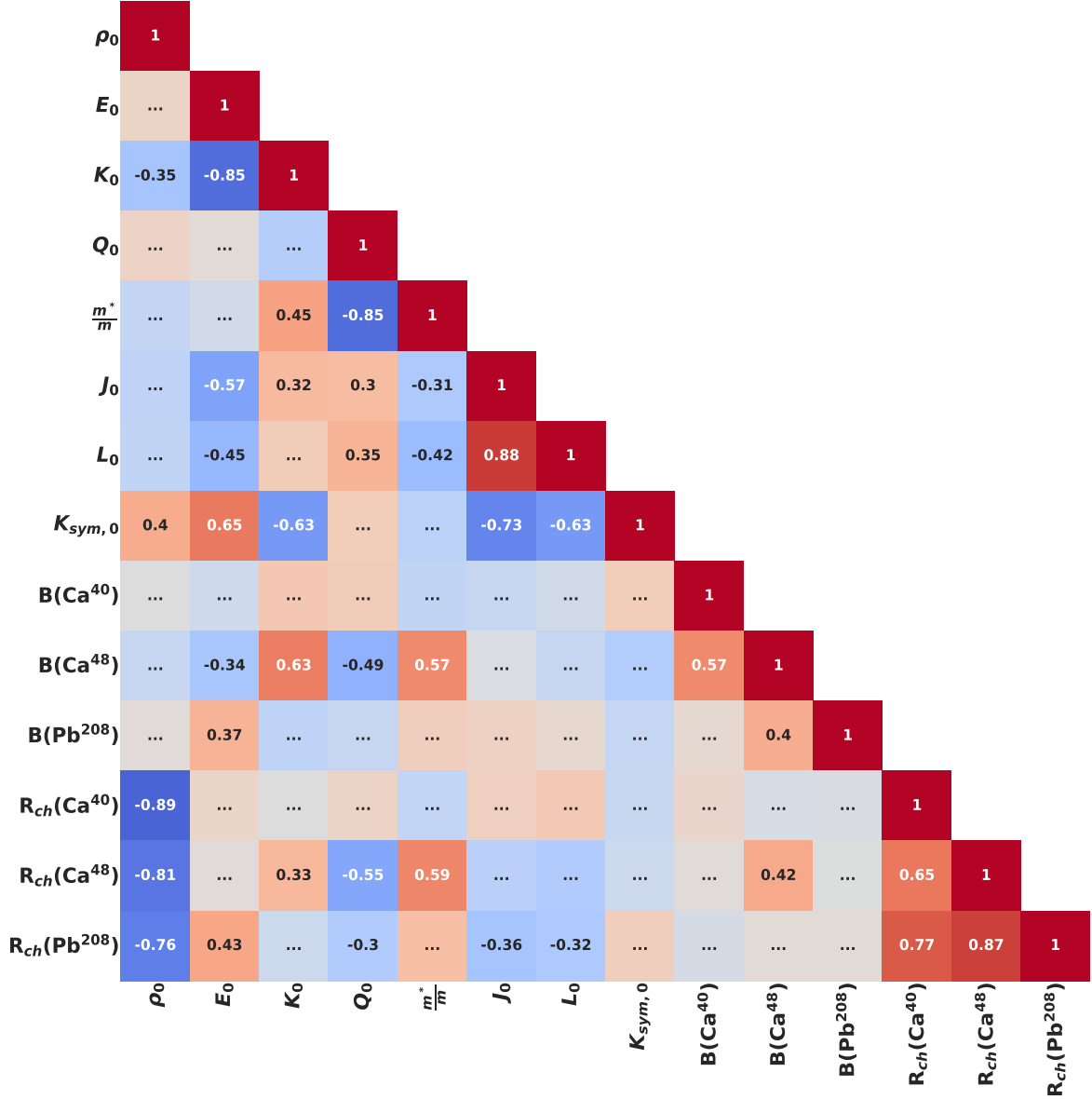}
    \caption{same as Fig. \ref{fig:corr_case1} but for the case of explicit FNC.}
    \label{fig:corr_case2}
\end{figure}
% \newpage
The Figs. \ref{fig:corr_case1} and \ref{fig:corr_case2} are the correlation heat maps showing the pairwise Pearson correlation coefficients ($r$) among the nuclear matter parameters and finite nuclei properties like binding energies and charge radii of Ca$^{40}$,Ca$^{48}$ and Pb$^{208}$. It is significantly noticeable that the correlations among the NMPs themselves are distinct for the cases of implicit and explicit FNCs, affecting the resulting distributions of the EOSs and hence the NS properties (as detailed in the main paper). 

 % The posterior distribution of the coupling parameters  $g_\sigma$, $g_\omega$, $g_\rho$, $A$, $B$, and $\Lambda_v$ is listed in Table \ref{coupling}.
% \newpage
\renewcommand{\thesection}{C}
\section{Results for model without the vector self-coupling} % ($C=0$)
In this section, we briefly outline some of the results for the RMF model without the vector self coupling term. This interaction is controlled by the coupling parameter $C$, which we take to be 0 for this section. 

Fig.\ref{fig:EOS}a shows the 95\% CI for the posterior distributions of EOSs resulting from the explicit and implicit FNCs. The two scenarios display distinct behaviors at different densities. At low densities (0.5 - $\sim$ 1.5 $\rho_0$), the EOSs generated for the case of explicit constraints are softer than those of the implicit constraints. As the density rises, the EOSs for the explicit FNC become stiffer due to the limits placed on the minimum-maximum mass. The values of KL divergence are shown in Fig.\ref{fig:EOS}b to vary with the density. These values indicate significant differences at low and high densities. However, at the densities $\sim 1.25\rho_0$, the noticeable overlap is reflected in the minimum value of the KL divergence. 
\renewcommand{\thefigure}{S3}
\begin{figure}[hb]
    \centering
    \includegraphics[width=0.75\textwidth,height=0.5\textwidth]{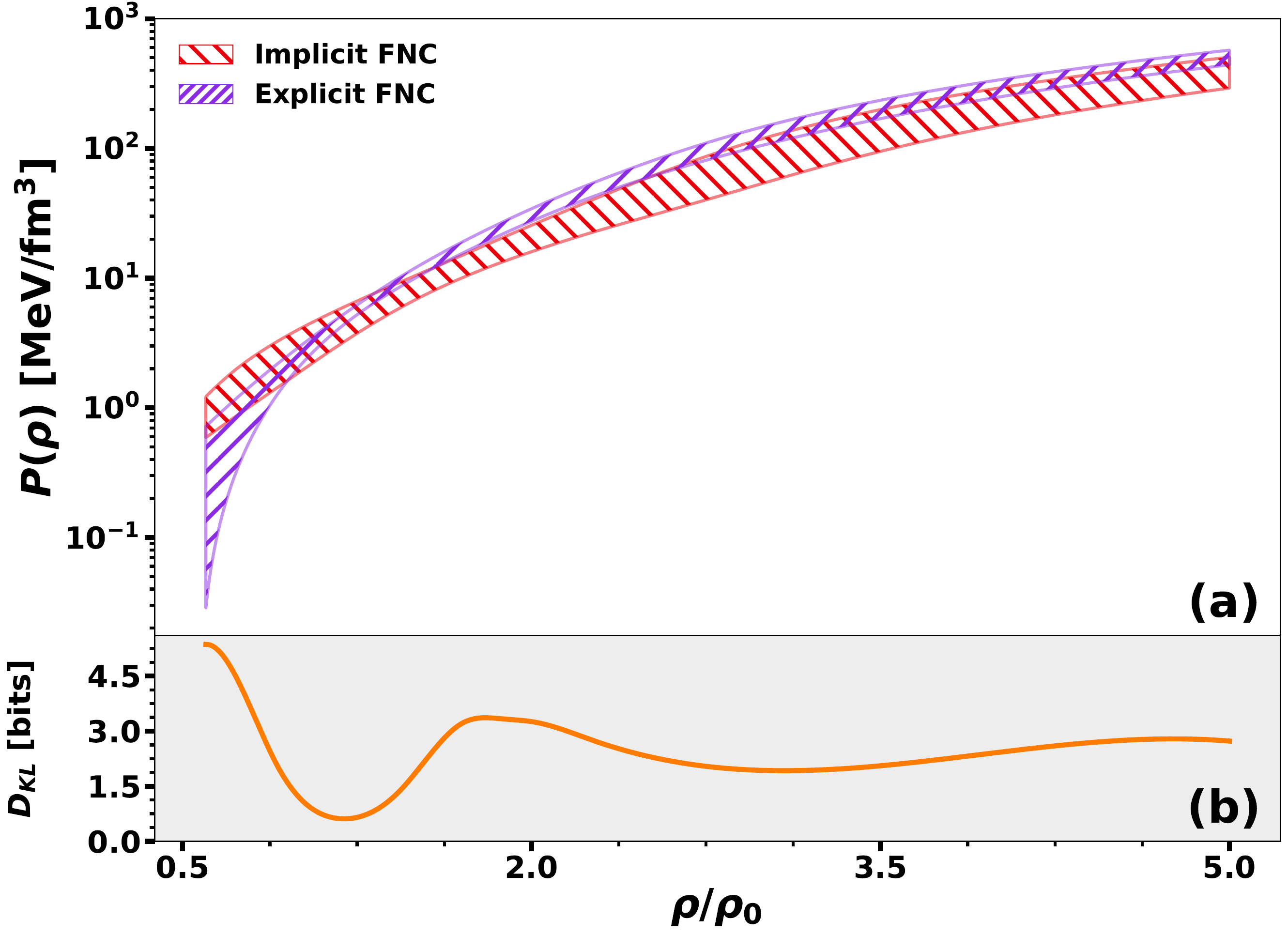}
    \caption{(a) EOSs obtained for the RMF model without the vector self-coupling ($C=0$). Shown here, the 95\% CI for the scenarios with implicit (red hatch) and explicit (violet hatch) finite nuclei constraints.(b) The Kullback-Leibler divergence (D$_{KL}$ in bits) was obtained using the distributions of EOSs for the two scenarios.}
    \label{fig:EOS}
\end{figure}
Table \ref{tab:liklelihood_ns_1} lists the properties of neutron stars, for the model without the vector self interaction ($C = 0$). It is observable that the radius for this case is considerably lower for intermediate mass ($\sim 1.44 M_\odot$) NSs by $\sim 0.5$ km, respectively as compared to the model with the quartic vector self interaction term. On the other hand, the radius of high mass NS ($\sim 2.08 M_\odot$) is smaller by the same amount. The uncertainty in the NS properties is also reduced for the model without the vector self interaction.
\renewcommand{\thetable}{S5}
\begin{table}[htb]
\begin{center}
    \caption{\label{tab:liklelihood_ns_1}% 
    Same as Table \ref{tab:liklelihood_ns} but obtained for the RMF model without the quartic vector self-coupling term (i.e. $C = 0$)}
    \bgroup
\def\arraystretch{1.5}
    \begin{tabular}{p{2.5cm}p{5cm}p{1.5cm}p{2cm}p{2cm}p{1cm}}
    \hline
    \hline
    Quantity& Source & Obs. & Implicit FNC & Explicit FNC & Ref. \\
    \hline
    % Maximum Mass & PSR J0740+6620 & $\geq 2.08 \pm 0.07$ & 2.28$^{+0.11}_{-0.10}$ & 2.49$^{+0.06}_{-0.05}$&\cite{fonseca2021refined}\\ 
    % \hline
    R (1.34 \msun) & NICER (PSR J0030+0451) Riley$^{*}$ & 12.71$^{+1.14}_{-1.19}$ & 12.59$^{+0.44}_{-0.39}$&12.78$^{+0.36}_{-0.26}$&\cite{riley2019nicer}\\
    R (1.44 \msun) & NICER (PSR J0030+0451) Miller$^{*}$ & 13.02$^{+1.24}_{-1.06}$ & 12.53$^{+0.44}_{-0.39}$&12.82$^{+0.33}_{-0.24}$&\cite{miller2019psr}\\
    R (2.07 \msun) & NICER (PSR J0740+6620) Riley$^{*}$ & 12.39$^{+1.30}_{-0.98}$ & 11.94$^{+0.49}_{-0.57}$&12.84$^{+0.2}_{-0.21}$&\cite{riley2021nicer}\\
    R (2.08 \msun) & NICER (PSR J0740+6620) Miller$^{*}$ & 13.7$^{+2.6}_{-1.5}$ &  11.92$^{+0.58}_{-0.50}$&12.84$^{+0.21}_{-0.21}$&\cite{miller2021radius}\\
    % \hline
    $\Lambda$ (1.4 \msun) & GW170817 &190$^{+390}_{-120}$&514.96$^{+85.51}_{-77.98}$&673.29$^{+73.48}_{-69.59}$&\cite{abbott2018gw170817rad}\\
    \hline
    \hline
    \end{tabular}
    \egroup
\end{center}
    \end{table}

% \newpage
\bibliographystyle{elsarticle-num}
\bibliography{references}